\documentclass[twocolumn,prl,superscriptaddress,preprintnumbers,amsmath,amssymb,longbibliography]{revtex4-1}
\usepackage{graphicx}
\usepackage{color}

\def\gtrless{\raise2.5pt\hbox{$>$}\llap{\lower2.5pt\hbox{$<$}}}
\def\gtrapprox{\raise2.5pt\hbox{$>$}\llap{\lower2.5pt\hbox{$\approx$}}}

\newcommand{\bsq}[1]{\begin{subequations}\label{#1}}
\newcommand{\esq}{\end{subequations}}
\newcommand{\beq}[1]{\begin{equation}\label{#1}}
\newcommand{\eeq}{\end{equation}}
\newcommand{\beqa}[1]{\begin{eqnarray}\label{#1}}
\newcommand{\eeqa}{\end{eqnarray}}
\newcommand{\add}[1]{#1}
\newcommand{\rem}[1]{}

\newcommand{\pP}{{\cal P}}

\newcommand{\rb}{{\bf r}}
\newcommand{\qb}{{\bf q}}

\newcommand{\vb}{{\bf v}}

\newcommand{\sigb}{\boldsymbol{\sigma}}

\renewcommand{\rho}{\varrho}
\renewcommand{\epsilon}{\varepsilon}

\begin{document}

\title{Emergence of long-ranged stress correlations  at the liquid to glass transition}

\author{Manuel Maier}
\affiliation{University of Konstanz, D-78457 Konstanz, Germany}
\author{Annette Zippelius}
\affiliation{University G\"ottingen, D-37077 G\"ottingen, Germany}
\author{Matthias Fuchs}
\affiliation{University of Konstanz, D-78457 Konstanz, Germany}

\date{\today}

\begin{abstract}

  A theory for the non-local shear stress correlations in supercooled
  liquids is derived from first principles. It captures the crossover
  from viscous to elastic dynamics at an idealized liquid to glass
  transition and explains the emergence of long-ranged stress
  correlations in glass, as expected from classical continuum
  elasticity. The long-ranged stress correlations can be traced to the
  coupling of shear stress to transverse momentum,
  which is ignored in the classic Maxwell model.
  To rescue this widely used model, we suggest a
  generalization in terms of a single relaxation time $\tau$ for the
  fast degrees of freedom only. This generalized Maxwell model implies
  a divergent correlation length $\xi\propto\tau$ as well as dynamic critical scaling and correctly
  accounts for the far-field stress correlations. It can be
  rephrased in terms of generalized hydrodynamic equations, which
  naturally couple stress and momentum and furthermore allow to
  connect to fluidity and elasto-plastic models.

\end{abstract}

\maketitle

In 1867, Maxwell described the phenomenology of viscoelasticity  in quiescent glass-forming liquids \cite{Maxwell}. A viscoelastic liquid possesses a slow structural process characterized by the (final) relaxation time $\tau$ and behaves like a solid with (shear) elastic modulus $G_\infty$ in rapid deformations.  Only finally, the liquid flows with a finite viscosity $\eta$, which follows as $\eta=G_\infty \tau$ according to Maxwell. The approach to vitrification is modeled by an increase in the relaxation time, and a glass state is probed when $\tau$ exceeds the observation time.  This model for the macroscopic stress response close to equilibrium has been the basis for phenomenological extensions to flowing complex fluids \cite{Larson}, yet it had been overlooked that it errs in the solid state.
Equilibrium correlations in fluids are short-ranged\add{, whereas} elastic stress correlations in solids\rem{ on the other hand} are long-ranged according to classical continuum mechanics \cite{Landau}\rem{, and the emergence of shear rigidity in glass is still a topic of active research \cite{Lemaitre2014,Szamel2015,Wittmer2015,Saw2016,Chowdhury2016}}. 
\add{Local stresses in amorphous solids have been a topic of
  strong interest in recent years, focusing on the rheology of
  viscoplastic materials~\cite{Sollich,Hebraud,Bocquet,Falk,Maloney}.
  The yielding of soft glassy materials is well acounted for in
  phenomenological models, which combine elastic deformations at small
  stresses with plastic deformations at large stresses. Well inside
  the glassy phase, the latter have been identified with activation
  processes between metastable states~\cite{Sollich,Hebraud,Bocquet},
  or with localised excitations, such as shear deformation
  zones~\cite{Falk} and quadrupolar energy
  fluctuations~\cite{Maloney}. Their coupling via long-ranged elastic stress fields has been incorporated in elasto-plastic models \cite{Picard2004,Ferrero2014,Nicolas2014}.
}

 \add{Here we focus on the liquid to glass transition and the emergence of
  shear rigidity \cite{Lemaitre2014,Szamel2015,Wittmer2015,Saw2016,Chowdhury2016}.} 
  Following Maxwell in assuming that correlations at finite
  frequencies cross over smoothly at a glass transition, the emergence
  of elasticity thus requires the build-up of long-lived and
  long-ranged spatial correlations in supercooled states.  We consider
  spatial correlations of the shear stress within the Zwanzig-Mori
  formalism \cite{Forster}, thereby generalize Maxwell's macroscopic
  description to finite wavevectors $\qb$. We recover the far-field
  solutions of elasticity theory in isotropic
  solids~\cite{Eshelby1957} in the limit of large $\tau$ \add{and identify
  its precursors in the fluid phase}. The generalized Maxwell model
  contains a correlation length $\xi$ diverging at the glass
  transition and possesses a non-analytic limit for small $\qb$ in
  glass.

We start from the conservation of total momentum in a fluid of $N$ particles with mass  $m$
which introduces the stress tensor $\sigb$ as the momentum current: $  \partial_t m \vb(\qb, t) = 
i \qb \cdot \sigb(\qb,t)$. The fluid velocity field is $\vb(\qb,t)=\frac{1}{\sqrt{N}}\;  \sum_{i=1}^N \; e^{ i \qb\cdot \rb_i(t)}\; \vb_i(t)$ \cite{Hansen}, 
where, $\rb_i(t)$ and $\vb_i(t)$ denote position and velocity of particle $i$. Using Newton's equation of motion, $m\dot \vb_i(t)  = {\bf F}_i(t)$, the stress tensor contains a kinetic term, which we will neglect, and a potential term, whose expression was determined by Irving and Kirkwood \cite{IrvingKirkwood}.  The potential term dominates in supercooled and glassy states \cite{Ladd1988}  and its qualitative change at the glass transition is the topic we want to address. Fluctuations are decomposed into plane-wave contributions at $\qb$ by Fourier-transformation.

The crucial difference between a fluid and a solid concerns the response under volume-conserving shear deformations; a fluid flows with a viscosity while a solid deforms dominantly elastically. 
{In both cases, the force transmitted by the stress through a planar element  is coplanar to it, which is captured by an off-diagonal element of the stress tensor \cite{Landau}. We chose $\sigma_{xy}(\qb)$,  and  consider its auto-correlation function:}
\beq{n2}
C_{\sigma}(\qb,t) = \frac{n}{k_BT} \; \langle \sigma_{xy}(\qb,t)^* \; 
 \sigma_{xy}(\qb) \rangle \; ,
 \eeq
 where any other set of orthogonal directions other than the $\hat{\bf x}$- and $\hat{\bf y}$-directions would be equivalent; here $n$ is the particle density and $k_BT$ the thermal energy.
 Also, we will consider the limit of an incompressible and isothermal fluid in order to simplify the presentation, postponing the compressible case of interest for the study of colloidal dispersions to a future submission \cite{Maier2017}.
The hydrodynamic conservation laws cause slow dynamics in {the shear stress auto-correlation function} 
$C_{\sigma}(\qb,t)$,
which can be brought out by projections onto the conserved variables according to the Zwanzig-Mori (ZM) formalism \cite{Forster}. 
 A projection operator $\pP$ captures the overlaps between general fluctuations and fluctuations of the (relevant)  conserved quantities, which are the transversal momenta in the present (incompressible, isothermal) case; $\pP=\frac{m}{k_BT}\, \vb^\perp(\qb) \rangle\!\cdot\! \langle \vb^\perp(\qb)^\ast$, with $ \vb^\perp(\qb)= \qb\!\times\!(\qb\!\times\!\vb(\qb))/q^2$. Non-local and non-Markovian effects follow from integrating out internal degrees of freedom and introduce memory kernels, which  reduce to transport coefficients in the hydrodynamic limit of  {both} small frequencies and {small} wavevectors \cite{Kadanoff}.
 G\"otze and Latz formulated a general ZM decomposition which is aimed at the limit of generalized hydrodynamics (GH) in viscoelastic fluids{, which retains the possibility of a large structural relaxation time $\tau$} \cite{Latz}. 
Applied to correlations of the shear stress, we find for the Laplace transform  of the stress auto-correlation 
\beq{o1}
C^{\rm GH}_{\sigma}(\qb,s) ={\rm G}^\perp_0(s)- \big((q_x^2+q_y^2)-4\frac{q_x^2q_y^2}{q^2} \big) \frac{({\rm G}^\perp_0(s))^2}{n k_BT}  \; {\rm K}^\perp_q(s)\;.
\eeq
(We have used the convention $C(s) = \int_0^\infty\!\!dt\, e^{-st} C(t)$ and only kept the leading order in $q$; for details of the derivation, see the supplementary material.)
The GH representation of Eq.~\eqref{o1} shows explicitly the decomposition of stress relaxation into a contribution of the hydrodynamic modes and the fast dynamics in the space complementary to the hydrodynamic modes. The latter are captured by the time- or frequency-dependent memory kernel ${\rm G}^\perp_{q=0}(s)$, whereas 
the former enter Eq.~\eqref{o1} via the  
transverse current correlation function:  $\langle \vb^\perp(\qb,t)^*\; \vb^\perp(\qb) \rangle = ({\bf 1} - \frac{\qb\qb}{q^2})\; {\rm K}^\perp_q(t)$. 
In the GH-limit, ${\rm K}^\perp_q(s)$ is conveniently represented \cite{Hansen}
\beq{mct2}
{\rm K}^{\perp}_q(s) =  \frac{k_BT/m}{s + \frac{q^2}{mn} {\rm G}^\perp_0(s)} 
\eeq
in terms of the same memory kernel ${\rm G}^\perp_0(s)$. The kernel is thus identified as generalised shear viscosity, called shear modulus in rheometry \cite{Larson},  which is the auto-correlation of the fluctuating transverse force with ZM-reduced dynamics \cite{Hansen,Forster}.

Equation~\eqref{o1} is the central result of our paper and will be discussed in the following for supercooled liquids, glasses and the transition in between. It is valid for arbitrary frequencies and contains
microscopic motion on short time scales. {A related function, the transverse force auto-correlation with real dynamics, which does not exhibit elastic correlations  \cite{Evans1981}, can be recovered from $C_{\sigma}(\qb,t)$ for specific wavevector directions (viz.~$\qb=q\hat{\bf x}$ and $\qb=q\hat{\bf y}$).} The crucial differences
between fluids and glasses are observed for small frequencies or long
timescales.  We hence focus on stress correlations on
hydrodynamic scales. This is also the regime, where the ZM
formalism develops its full power because the relaxation of selected
variables can be well separated from microscopic scales. {We already used this in Eq.~\eqref{mct2}, where the hydrodynamic pole in the (conserved) momentum fluctuations was identified, which captures shear diffusion in the limit of small frequencies and wavevectors \cite{Forster,Hansen}.}

For the global shear stress in fluid states,  the difference between ZM-projected  and full dynamics vanishes as expected  \cite{Forster}:
 \beq{o3}
 C_{\sigma}(\qb={\bf 0},t)= {\rm G}^\perp_{0}(t)\;,\; \mbox{(fluid)} \;. 
 \eeq
The global memory kernel reduces to the shear viscosity for vanishing frequency, so Eq.~\eqref{o3} is equivalent to the Green-Kubo relation for the viscosity \cite{Hansen}:
\beq{o4}
 \eta=  {\rm G}^\perp_{q=0}(s=0) = \frac{n}{k_BT} \int_0^\infty\!\!\! \!\! dt \, \langle \sigma_{xy}(\qb\!=\!{\bf 0},t)  
 \sigma_{xy}(\qb\!=\!{\bf 0}) \rangle 
 \eeq
The hydrodynamics of fluids is recovered by replacing the memory
kernels by wavevector and frequency independent transport
coefficients. {Here this means ${\rm G}^\perp_{q=0}(s=0)\to\eta$,} and gives for the shear stress correlation
function in a liquid:
   \beq{o6}
C_{\sigma}^{\rm fluid}(\qb,s) =\eta +
\big(\frac{q_x^2q_y^2}{q^2} -\frac{q_x^2+q_y^2}{4} \big) \frac{4\,\eta^2}{nm s + \eta q^2} \eeq
Clearly, the hydrodynamic velocity correlator introduces {into the stress correlator a small correction to the viscosity; it is of order $q^2$ and anisotropic.}

Maxwell`s model of viscoelasticity introduces a single relaxation time $\tau$ for the stress, so that its correlation is given by 
$C^{\rm Max}_{\sigma}(\qb,s) =\frac{G_\infty \tau}{1+s\tau}$. {(Maxwell ignored the wavevector dependence.)}
Solid-like behavior emerges, when the deformations are rapid relative to $\tau$ and elastic correlations dominate at low frequencies, implying $\lim_{s\to0}\lim_{\tau \to \infty} sC^{\rm Max}_{\sigma}(\qb,s)=G_\infty$. As function of time, this implies a persistent contribution:
$\lim_{t\to\infty}\lim_{\tau \to \infty} C^{\rm Max}_{\sigma}(\qb,t)=G_\infty$. For the generalized hydrodynamics, entailed in the full Eq.~\eqref{o1}, we still expect a divergence of the relaxation time $\tau$, however the small q-dependence is completely different {than envisioned by Maxwell} due to the coupling to correlations of transverse momentum which are long ranged in an elastic solid.
{To elucidate this we follow  the suggestion} by the microscopic mode-coupling theory that an idealized glass state is obtained when the
relaxation time $\tau$ is infinite and correlation
functions do not decay to zero \cite{Goetze}.  This
should hold for collective density fluctuations and the fluctuating
force memory kernel ${\rm G}^\perp_{0}(t)$ \cite{Leutheusser}, implying a time-persistent contribution to the shear stress correlation:
 \bsq{o9}
  \beq{o9a}
 C_{\sigma}(\qb,t) \to  C_{\sigma}^\infty(\qb) \,,\; \mbox{for } t\to\infty\;\mbox{ in glass} ,
 \eeq
In the GH-limit of an incompressible glass, it  is solely determined by ${\rm G}^\perp_{0,\infty}=\lim_{t\to\infty}{\rm G}^\perp_{0}(t)$ and  follows from Eq.~\eqref{o1}:
 \beq{o9b}
 C_{\sigma}^\infty(\qb) =  \big( 4 \frac{q_x^2q_y^2}{q^4}\!+\!\frac{q_z^2}{q^2}\big) \;  {\rm G}^\perp_{0,\infty}
 +{\cal O}(q^2) 
 \eeq
The system is characterized by a finite resistivity to shear deformations and Maxwell's shear elastic constant can be identified with $G_\infty={\rm G}^\perp_{0,\infty}$. In contrast to Maxwell`s macroscopic description of a glass, {however,}  generalised hydrodynamics predicts that the  limit of $q\to 0$ is non-analytic and depends on the direction $\qb$ is taken to zero. 
 Such nonanalytic behaviour points to the existence of a length
 scale that diverges in the supercooled regime when approaching the glass transition. It will be discussed below.
Eshelby's result for  the response of an elastic (isotropic) medium to a point force  can be recognized  in Eq.~\eqref{o9b} \cite{Eshelby1957,Picard2004}. It predicts the existence of  long-ranged stress correlations:
 \beq{o10}
C_{\sigma}^\infty(\rb) \to \frac{3}{4\pi}\, \frac{{\rm G}^\perp_{0,\infty}}{r^3} \;  \frac{10x^2y^2-r^2(x^2+y^2)}{r^4}
  \;,\quad \mbox{for } r\to\infty
 \eeq \esq
In $d$ dimensions, the elastic correlations decay like $r^{-d}$, as seen recently in flowing dense quasi-two-dimensional emulsions \cite{Desmond2015}. 

How can we generalise Maxwell`s theory to account for the
frequency and wavevector dependence of stress correlations at the
glass transition and in the elastic solid? Generalised hydrodynamics, as 
entailed in Eq.~\eqref{o1}, suggests to approximate the memory kernel instead of the correlations of the stress itself. Thus 
 we model the memory kernel by a single relaxation time $\tau$
 \bsq{oo30}
\beq{o30}{\rm G}^\perp_{0}(s) \approx {\rm G}^{\rm \add{g}M}(s) = \frac{G_\infty \tau}{1+s\tau} \;.
\eeq
in the spirit of Maxwell but correctly accounting for the coupling of the stress to conserved momentum fluctuations. Substitution into Eqs.~\eqref{o1}  then captures the far-field shear stress correlation function in the incompressible limit to be denoted $C_{\sigma}^{\rm gM}(\qb,s)$.
The generalised 
  Maxwell model has a rather rich content. The angular
dependence contains an isotropic term and (choosing standard spherical
coordinates) the dependence
$P_{xy}( \vartheta,\varphi) = 4\hat{q}_x^2 \hat{q}_y^2 + \hat{q}_z^2 =
\frac 12 ( 1 +\cos^4{\vartheta} - \cos{4\varphi} \sin^4{\vartheta}$),
which agrees with the one of the elastic Green's function in Eq.~\eqref{o9b}. 
The distance to the idealised glass transition is controlled by the
divergence of $\tau$ which we do not specify explicitly.
Introducing a characteristic lengthscale $\xi^2= \frac{G_\infty\tau^2}{mn}= v_T^2 \tau^2$ (with $v_T$ the transversal sound velocity of glass \cite{Goetze,Ahluwalia1998}), we observe that the time and wavenumber dependent stress correlation obeys scaling:
\beq{o31}
C_{\sigma}^{\rm gM}(\qb,t) = {\cal F}( t/\tau , q\xi , P(\vartheta,\varphi)) \;,
\eeq
\esq
i.e.~all dependence on the distance to the critical point is absorbed in the timescale $\tau$ and the length scale $\xi$.
The correlation length  $\xi\propto\tau$ determines the spatial extent of solid like regions within the viscoelastic fluid. It diverges strongly as the glass transition is approached signaling the appearance of long-ranged stress correlations in the glassy state. 
The hydrodynamic excitations are determined by the poles of
$C_{\sigma}^{\rm gM}(\qb,s)$ in the complex s-plane:
$s\tau(1+s\tau)+q^2\xi^2=0$. The hydrodynamics of the fluid is
recovered in the limit $s\tau\ll 1$, implying
$s\tau=-q^2\xi^2$.
 Glassy hydrodynamics is recovered in
the opposite limit $s\tau\gg 1$, implying $s\tau=\pm iq\xi$. The
critical dynamics is contained in the roots of the dispersion
relation
for $q\xi\gg 1$. In this limit, one finds again $s\tau=\pm iq\xi$,
i.e.~small wavelength, high frequency sound in the fluid phase \cite{Goetze,Ahluwalia1998}.  Adding an instantaneous dissipation rate to Eq.~\eqref{o30} is an easy way to capture sound damping \cite{Maier2017}.

The stress correlation tensor can be measured in a linear response experiment by applying a weak shear velocity gradient $\partial_x {\rm v}_y^{\rm ext}$ to the liquid \cite{Forster}.
The inhomogeneous flow gives rise to a shear stress which is given in linear response by
\beq{lr4} 
   \langle \sigma_{xy}(\rb,t) \rangle^{\rm lr}  
=  2  \int_{-\infty}^t\!\!\!\!\!\! dt'  \int\!\! d^dr'\; C_{\sigma}(\rb-{\bf r'},t-t') \; \bar{\kappa}^{\rm ext}_{xy}({\bf r'},t').
\eeq
Because of the symmetry of the Irving-Kirkwood stress tensor,  only the symmetric component of the external velocity gradient enters, with 
$\bar{\kappa}_{\alpha\beta}(\rb,t)= \frac 12 \left( \partial_\alpha v_\beta(\rb,t)+\partial_\beta v_\alpha(\rb,t) \right)$. 
In the fluid phase, a uniform stationary shear flow gives rise to
uniform stationary stresses \cite{Hansen}:
$\sigma_{xy}=2 \eta \,\bar{\kappa}^{\rm ext}_{xy}$. This global
constitutive equation is generalised by Eq.~\eqref{o1} to finite
wavevectors and frequencies,
$\sigma_{xy}(\qb,\omega)=2 C_{\sigma}(\qb,s=-i\omega)\,
\bar{\kappa}^{\rm ext}_{xy}(\qb,\omega)$.  Here, a periodic external flow rate with frequency
$\omega$ was assumed, which leads to a periodic stress with the same
frequency. 

The linear response relation allows for an intuitive interpretation of generalised hydrodynamics. Momentum conservation is expressed as usual in terms of the linearized Navier-Stokes equation with the local pressure $p$ 
\bsq{lr5}
\beq{lr5b}
m n \partial_t \vb(\rb,t) = \nabla \cdot\left( \sigb^{\rm gM}(\rb,t) - p(\rb,t) {\bf 1} \right)
\eeq
We consider the incompressible limit (viz.~\mbox{$\nabla \cdot {\vb}=0$}), so that the velocity is purely transverse $\vb^\perp$ and does not couple to the pressure. The constitutive equation for $\sigb^{\rm gM}(\rb,t)$
\beq{lr5a}
\left( \frac{1}{\tau} + \partial_t \right) \sigb^{\rm gM}(\rb,t) = 2 G_\infty \left( \bar{\boldsymbol\kappa}^{\rm ext}(\rb,t) + \bar{\boldsymbol \kappa}(\rb,t) \right)\;
\eeq
\esq
is built on Maxwell`s insight on glassy relaxation but includes the full velocity gradient, which is the sum of the externally applied one and the internal flow as computed from Eq.~\eqref{lr5b}, resulting in a linear but spatially 
nonlocal differential equation. Fourier transformation of 
Eqs.~\eqref{lr5b} and ~\eqref{lr5a} reproduces the linear response relation 
with $C_{\sigma}^{\rm gM}(\qb,s)$ as given by Eq.~\eqref{o1} with the Maxwell approximation for ${\rm G}^\perp_{0}(s)$.

\rem{While the}\add{The} classic Maxwell model \rem{would only have}\add{only has} the external flow gradient on the right hand side in Eq.~\eqref{lr5a}. \add{In the generalized model,} the internal $\bar{\boldsymbol \kappa}$ arises from the non-local velocity field which necessarily \rem{(because of hydrodynamic flow in Eq.~\eqref{lr5b})} is induced by the imposed flow \rem{also}\add{(Eq.~\ref{lr5b})}. Only for finite frequencies and relaxation times $\tau$, does the transverse momentum diffusion allow for an (anisotropic) gradient expansion leading to Eq.~\eqref{o6}. In solid states where $1/\tau=0$,  however, the gradient expansion breaks down and the non-local strain field induces the non-analytic small $\qb$-expansion in Eq.~\eqref{o9}, which signals long-ranged elastic stress fields that are at the heart of \add{the} elasto-plastic models \cite{Picard2004,Ferrero2014,Nicolas2014}. The strain field considered in these models follows here as the time integral of the (symmetrized) velocity gradient tensor \cite{Goldhirsch2002}: $\boldsymbol{\epsilon}(\rb,t) 
=  \int_{t_0}^{t_0+t} dt' \; \bar{\boldsymbol\kappa}(\rb,t')$.

In summary, within the Zwanzig-Mori approach we have obtained the non-local correlations of the shear stress in the long-wavelength limit. \rem{While this}\add{This} result holds generally in viscoelastic liquids including e.g.~\rem{in }polymeric systems.\rem{ we}\add{We specifically} addressed glass-forming  melts. We have shown that the most simple generalization of Maxwell's model including spatial variations of the  stress recovers the long-ranged elastic fields expected in  solids. The shear-stress memory kernel  plays the role attributed by Maxwell to the global shear stress. The generalization implies the rapid growth of a correlation length $\xi$, which opens the window in wavevector space for the non-analytic small-$\qb$ behavior of the shear stress auto-correlator expected in solids. The far-field decay of the frozen-in stress fluctuations $C_\sigma^\infty(\rb) \propto r^{-d}$ agrees with the one deduced from Goldstone modes in solids with quenched disorder in $d=2$ and $d=3$ \cite{Mukhopadahyay2004}.
 Our approach to neglect the wavevector dependence of the generalized viscosity kernel ${\rm G}^\perp_q(t)$ is at odds with some simulation results  {\cite{Furukawa2011,Puscasu2010}} which appear to find a strong wavevector dependence of the viscosity when supercooling. {(Whether this can be related  to the non-analytic $\qb$-dependent stress correlations we find, should be clarified in future.)} Yet, our approach may be useful for non-local rheological models, where ad-hoc transport equations are formulated including diffusive terms \cite{Olmsted2008} or considering the inverse  Maxwell relaxation time as independent state variable \cite{Picard2002,Goyon2010,Bouzid2015}. The generalized Maxwell model Eq.~\eqref{lr5} implies that  non-locality of the stress relaxation is transported with the velocity field accompanying the externally imposed flow, which constrains the model-building.

\begin{acknowledgments}

We thank J. Baschnagel for discussions and
acknowledge support from the DFG through FOR 1394 projects P3 and P6.
\end{acknowledgments}

\bibliography{lit}

\begin{thebibliography}{37}%
\makeatletter
\providecommand \@ifxundefined [1]{%
 \@ifx{#1\undefined}
}%
\providecommand \@ifnum [1]{%
 \ifnum #1\expandafter \@firstoftwo
 \else \expandafter \@secondoftwo
 \fi
}%
\providecommand \@ifx [1]{%
 \ifx #1\expandafter \@firstoftwo
 \else \expandafter \@secondoftwo
 \fi
}%
\providecommand \natexlab [1]{#1}%
\providecommand \enquote  [1]{``#1''}%
\providecommand \bibnamefont  [1]{#1}%
\providecommand \bibfnamefont [1]{#1}%
\providecommand \citenamefont [1]{#1}%
\providecommand \href@noop [0]{\@secondoftwo}%
\providecommand \href [0]{\begingroup \@sanitize@url \@href}%
\providecommand \@href[1]{\@@startlink{#1}\@@href}%
\providecommand \@@href[1]{\endgroup#1\@@endlink}%
\providecommand \@sanitize@url [0]{\catcode `\\12\catcode `\$12\catcode
  `\&12\catcode `\#12\catcode `\^12\catcode `\_12\catcode `\%12\relax}%
\providecommand \@@startlink[1]{}%
\providecommand \@@endlink[0]{}%
\providecommand \url  [0]{\begingroup\@sanitize@url \@url }%
\providecommand \@url [1]{\endgroup\@href {#1}{\urlprefix }}%
\providecommand \urlprefix  [0]{URL }%
\providecommand \Eprint [0]{\href }%
\providecommand \doibase [0]{http://dx.doi.org/}%
\providecommand \selectlanguage [0]{\@gobble}%
\providecommand \bibinfo  [0]{\@secondoftwo}%
\providecommand \bibfield  [0]{\@secondoftwo}%
\providecommand \translation [1]{[#1]}%
\providecommand \BibitemOpen [0]{}%
\providecommand \bibitemStop [0]{}%
\providecommand \bibitemNoStop [0]{.\EOS\space}%
\providecommand \EOS [0]{\spacefactor3000\relax}%
\providecommand \BibitemShut  [1]{\csname bibitem#1\endcsname}%
\let\auto@bib@innerbib\@empty
\bibitem [{\citenamefont {Maxwell}(1867)}]{Maxwell}%
  \BibitemOpen
  \bibfield  {author} {\bibinfo {author} {\bibfnamefont {J.~C.}\ \bibnamefont
  {Maxwell}},\ }\bibfield  {title} {\enquote {\bibinfo {title} {On the
  dynamical theory of gases},}\ }\href@noop {} {\bibfield  {journal} {\bibinfo
  {journal} {Philos. Trans. R. Soc. A}\ }\textbf {\bibinfo {volume} {157}},\
  \bibinfo {pages} {49} (\bibinfo {year} {1867})}\BibitemShut {NoStop}%
\bibitem [{\citenamefont {Larson}(1999)}]{Larson}%
  \BibitemOpen
  \bibfield  {author} {\bibinfo {author} {\bibfnamefont {R.~G.}\ \bibnamefont
  {Larson}},\ }\href@noop {} {\emph {\bibinfo {title} {{The structure and
  rheology of complex fluids}}}}\ (\bibinfo  {publisher} {Oxford University
  Press},\ \bibinfo {address} {New York},\ \bibinfo {year} {1999})\BibitemShut
  {NoStop}%
\bibitem [{\citenamefont {Landau}\ \emph {et~al.}(1986)\citenamefont {Landau},
  \citenamefont {Pitaevskii}, \citenamefont {Lifshitz},\ and\ \citenamefont
  {Kosevich}}]{Landau}%
  \BibitemOpen
  \bibfield  {author} {\bibinfo {author} {\bibfnamefont {L.~D.}\ \bibnamefont
  {Landau}}, \bibinfo {author} {\bibfnamefont {L.~P.}\ \bibnamefont
  {Pitaevskii}}, \bibinfo {author} {\bibfnamefont {E.~M.}\ \bibnamefont
  {Lifshitz}}, \ and\ \bibinfo {author} {\bibfnamefont {A.~M.}\ \bibnamefont
  {Kosevich}},\ }\href@noop {} {\emph {\bibinfo {title} {Theory of
  Elasticity}}}\ (\bibinfo  {publisher} {Butterworth-Heinemann},\ \bibinfo
  {year} {1986})\BibitemShut {NoStop}%
\bibitem [{\citenamefont {Sollich}\ \emph {et~al.}(1997)\citenamefont
  {Sollich}, \citenamefont {Lequeux}, \citenamefont {Hebraud},\ and\
  \citenamefont {Cates}}]{Sollich}%
  \BibitemOpen
  \bibfield  {author} {\bibinfo {author} {\bibfnamefont {P.}~\bibnamefont
  {Sollich}}, \bibinfo {author} {\bibfnamefont {F.}~\bibnamefont {Lequeux}},
  \bibinfo {author} {\bibfnamefont {P.}~\bibnamefont {Hebraud}}, \ and\
  \bibinfo {author} {\bibfnamefont {M.~E.}\ \bibnamefont {Cates}},\ }\bibfield
  {title} {\enquote {\bibinfo {title} {Rheology of soft glassy materials},}\
  }\href@noop {} {\bibfield  {journal} {\bibinfo  {journal} {Phys. Rev. Lett.}\
  }\textbf {\bibinfo {volume} {78}},\ \bibinfo {pages} {2020} (\bibinfo {year}
  {1997})}\BibitemShut {NoStop}%
\bibitem [{\citenamefont {Hebraud}\ and\ \citenamefont
  {Lequeux}(1998)}]{Hebraud}%
  \BibitemOpen
  \bibfield  {author} {\bibinfo {author} {\bibfnamefont {P.}~\bibnamefont
  {Hebraud}}\ and\ \bibinfo {author} {\bibfnamefont {F.}~\bibnamefont
  {Lequeux}},\ }\bibfield  {title} {\enquote {\bibinfo {title}
  {Mode-coupling-theory for the pasty rheology of soft glassy materials},}\
  }\href@noop {} {\bibfield  {journal} {\bibinfo  {journal} {Phys. Rev. Lett.}\
  }\textbf {\bibinfo {volume} {81}},\ \bibinfo {pages} {2934} (\bibinfo {year}
  {1998})}\BibitemShut {NoStop}%
\bibitem [{\citenamefont {Bocquet}\ \emph {et~al.}(2009)\citenamefont
  {Bocquet}, \citenamefont {Colin},\ and\ \citenamefont {Ajdari}}]{Bocquet}%
  \BibitemOpen
  \bibfield  {author} {\bibinfo {author} {\bibfnamefont {L.}~\bibnamefont
  {Bocquet}}, \bibinfo {author} {\bibfnamefont {A.}~\bibnamefont {Colin}}, \
  and\ \bibinfo {author} {\bibfnamefont {A.}~\bibnamefont {Ajdari}},\
  }\bibfield  {title} {\enquote {\bibinfo {title} {Kinetic theory of plastic
  flow in soft glassy materials},}\ }\href@noop {} {\bibfield  {journal}
  {\bibinfo  {journal} {Phys. Rev. Lett.}\ }\textbf {\bibinfo {volume} {103}}
  (\bibinfo {year} {2009})}\BibitemShut {NoStop}%
\bibitem [{\citenamefont {Falk}\ and\ \citenamefont {Langer}(1998)}]{Falk}%
  \BibitemOpen
  \bibfield  {author} {\bibinfo {author} {\bibfnamefont {M.L.}\ \bibnamefont
  {Falk}}\ and\ \bibinfo {author} {\bibfnamefont {J.S.}\ \bibnamefont
  {Langer}},\ }\bibfield  {title} {\enquote {\bibinfo {title} {Dynamics of
  viscoplastic deformation in amorphous solids},}\ }\href@noop {} {\bibfield
  {journal} {\bibinfo  {journal} {Phys. Rev. E}\ }\textbf {\bibinfo {volume}
  {57}} (\bibinfo {year} {1998})}\BibitemShut {NoStop}%
\bibitem [{\citenamefont {Maloney}\ and\ \citenamefont
  {Lemaitre}(2004)}]{Maloney}%
  \BibitemOpen
  \bibfield  {author} {\bibinfo {author} {\bibfnamefont {C.}~\bibnamefont
  {Maloney}}\ and\ \bibinfo {author} {\bibfnamefont {A.}~\bibnamefont
  {Lemaitre}},\ }\bibfield  {title} {\enquote {\bibinfo {title} {Subextensive
  scaling in the athermal, quasistatic limit of amorphous matter in plastic
  shear flow},}\ }\href@noop {} {\bibfield  {journal} {\bibinfo  {journal}
  {Phys. Rev. Lett.}\ }\textbf {\bibinfo {volume} {93}} (\bibinfo {year}
  {2004})}\BibitemShut {NoStop}%
\bibitem [{\citenamefont {Picard}\ \emph {et~al.}(2004)\citenamefont {Picard},
  \citenamefont {Ajdari}, \citenamefont {Lequeux},\ and\ \citenamefont
  {Bocquet}}]{Picard2004}%
  \BibitemOpen
  \bibfield  {author} {\bibinfo {author} {\bibfnamefont {G.}~\bibnamefont
  {Picard}}, \bibinfo {author} {\bibfnamefont {A.}~\bibnamefont {Ajdari}},
  \bibinfo {author} {\bibfnamefont {F.}~\bibnamefont {Lequeux}}, \ and\
  \bibinfo {author} {\bibfnamefont {L.}~\bibnamefont {Bocquet}},\ }\bibfield
  {title} {\enquote {\bibinfo {title} {Elastic consequences of a single plastic
  event: A step towards the microscopic modeling of the flow of yield stress
  fluids},}\ }\href@noop {} {\bibfield  {journal} {\bibinfo  {journal} {Eur.
  Phys. Jour. E}\ }\textbf {\bibinfo {volume} {15}},\ \bibinfo {pages} {371}
  (\bibinfo {year} {2004})}\BibitemShut {NoStop}%
\bibitem [{\citenamefont {Ferrero}\ \emph {et~al.}(2014)\citenamefont
  {Ferrero}, \citenamefont {Martens},\ and\ \citenamefont
  {Barrat}}]{Ferrero2014}%
  \BibitemOpen
  \bibfield  {author} {\bibinfo {author} {\bibfnamefont {E.~E.}\ \bibnamefont
  {Ferrero}}, \bibinfo {author} {\bibfnamefont {K.}~\bibnamefont {Martens}}, \
  and\ \bibinfo {author} {\bibfnamefont {J.-L.}\ \bibnamefont {Barrat}},\
  }\bibfield  {title} {\enquote {\bibinfo {title} {Relaxation in yield systems
  through elastically interacting activated events},}\ }\href@noop {}
  {\bibfield  {journal} {\bibinfo  {journal} {Phys. Rev. Lett.}\ }\textbf
  {\bibinfo {volume} {113}},\ \bibinfo {pages} {248301} (\bibinfo {year}
  {2014})}\BibitemShut {NoStop}%
\bibitem [{\citenamefont {Nicolas}\ \emph {et~al.}(2014)\citenamefont
  {Nicolas}, \citenamefont {Rottler},\ and\ \citenamefont
  {Barrat}}]{Nicolas2014}%
  \BibitemOpen
  \bibfield  {author} {\bibinfo {author} {\bibfnamefont {A.}~\bibnamefont
  {Nicolas}}, \bibinfo {author} {\bibfnamefont {J.}~\bibnamefont {Rottler}}, \
  and\ \bibinfo {author} {\bibfnamefont {J.-L.}\ \bibnamefont {Barrat}},\
  }\bibfield  {title} {\enquote {\bibinfo {title} {Spatiotemporal correlations
  between plastic events in the shear flow of athermal amorphous solids},}\
  }\href@noop {} {\bibfield  {journal} {\bibinfo  {journal} {Euro. Phys. J. E}\
  }\textbf {\bibinfo {volume} {37}},\ \bibinfo {pages} {50} (\bibinfo {year}
  {2014})}\BibitemShut {NoStop}%
\bibitem [{\citenamefont {Lema\^{\i}tre}(2014)}]{Lemaitre2014}%
  \BibitemOpen
  \bibfield  {author} {\bibinfo {author} {\bibfnamefont {A.}~\bibnamefont
  {Lema\^{\i}tre}},\ }\bibfield  {title} {\enquote {\bibinfo {title}
  {Structural relaxation is a scale-free process},}\ }\href@noop {} {\bibfield
  {journal} {\bibinfo  {journal} {Phys. Rev. Lett.}\ }\textbf {\bibinfo
  {volume} {113}},\ \bibinfo {pages} {245702} (\bibinfo {year}
  {2014})}\BibitemShut {NoStop}%
\bibitem [{\citenamefont {Flenner}\ and\ \citenamefont
  {Szamel}(2015)}]{Szamel2015}%
  \BibitemOpen
  \bibfield  {author} {\bibinfo {author} {\bibfnamefont {E.}~\bibnamefont
  {Flenner}}\ and\ \bibinfo {author} {\bibfnamefont {G.}~\bibnamefont
  {Szamel}},\ }\bibfield  {title} {\enquote {\bibinfo {title} {Long-range
  spatial correlations of particle displacements and the emergence of
  elasticity},}\ }\href@noop {} {\bibfield  {journal} {\bibinfo  {journal}
  {Phys. Rev. Lett.}\ }\textbf {\bibinfo {volume} {114}},\ \bibinfo {pages}
  {025501} (\bibinfo {year} {2015})}\BibitemShut {NoStop}%
\bibitem [{\citenamefont {Wittmer}\ \emph {et~al.}(2015)\citenamefont
  {Wittmer}, \citenamefont {Xu},\ and\ \citenamefont
  {Baschnagel}}]{Wittmer2015}%
  \BibitemOpen
  \bibfield  {author} {\bibinfo {author} {\bibfnamefont {J.~P.}\ \bibnamefont
  {Wittmer}}, \bibinfo {author} {\bibfnamefont {H.}~\bibnamefont {Xu}}, \ and\
  \bibinfo {author} {\bibfnamefont {J.}~\bibnamefont {Baschnagel}},\ }\bibfield
   {title} {\enquote {\bibinfo {title} {Shear-stress relaxation and ensemble
  transformation of shear-stress autocorrelation functions},}\ }\href@noop {}
  {\bibfield  {journal} {\bibinfo  {journal} {Phys. Rev. E}\ }\textbf {\bibinfo
  {volume} {91}},\ \bibinfo {pages} {022107} (\bibinfo {year}
  {2015})}\BibitemShut {NoStop}%
\bibitem [{\citenamefont {Saw}\ and\ \citenamefont
  {Harrowell}(2016)}]{Saw2016}%
  \BibitemOpen
  \bibfield  {author} {\bibinfo {author} {\bibfnamefont {S.}~\bibnamefont
  {Saw}}\ and\ \bibinfo {author} {\bibfnamefont {P.}~\bibnamefont
  {Harrowell}},\ }\bibfield  {title} {\enquote {\bibinfo {title} {Rigidity in
  condensed matter and its origin in configurational constraint},}\ }\href
  {\doibase 10.1103/PhysRevLett.116.137801} {\bibfield  {journal} {\bibinfo
  {journal} {Phys. Rev. Lett.}\ }\textbf {\bibinfo {volume} {116}},\ \bibinfo
  {pages} {137801} (\bibinfo {year} {2016})}\BibitemShut {NoStop}%
\bibitem [{\citenamefont {Chowdhury}\ \emph {et~al.}(2016)\citenamefont
  {Chowdhury}, \citenamefont {Abraham}, \citenamefont {Hudson},\ and\
  \citenamefont {Harrowell}}]{Chowdhury2016}%
  \BibitemOpen
  \bibfield  {author} {\bibinfo {author} {\bibfnamefont {S.}~\bibnamefont
  {Chowdhury}}, \bibinfo {author} {\bibfnamefont {S.}~\bibnamefont {Abraham}},
  \bibinfo {author} {\bibfnamefont {T.}~\bibnamefont {Hudson}}, \ and\ \bibinfo
  {author} {\bibfnamefont {P.}~\bibnamefont {Harrowell}},\ }\bibfield  {title}
  {\enquote {\bibinfo {title} {Long range stress correlations in the inherent
  structures of liquids at rest},}\ }\href@noop {} {\bibfield  {journal}
  {\bibinfo  {journal} {J. Chem. Phys.}\ }\textbf {\bibinfo {volume} {144}},\
  \bibinfo {pages} {124508} (\bibinfo {year} {2016})}\BibitemShut {NoStop}%
\bibitem [{\citenamefont {Forster}(1995)}]{Forster}%
  \BibitemOpen
  \bibfield  {author} {\bibinfo {author} {\bibfnamefont {D.}~\bibnamefont
  {Forster}},\ }\href {https://books.google.co.in/books?id=tdg6AAAACAAJ} {\emph
  {\bibinfo {title} {Hydrodynamic Fluctuations, Broken Symmetry, and
  Correlation Functions}}},\ Advanced book classics\ (\bibinfo  {publisher}
  {Perseus Books},\ \bibinfo {year} {1995})\BibitemShut {NoStop}%
\bibitem [{\citenamefont {Eshelby}(1957)}]{Eshelby1957}%
  \BibitemOpen
  \bibfield  {author} {\bibinfo {author} {\bibfnamefont {J.D.}\ \bibnamefont
  {Eshelby}},\ }\bibfield  {title} {\enquote {\bibinfo {title} {The
  determination of the elastic field of an ellipsoidal inclusion, and related
  problems},}\ }\href@noop {} {\bibfield  {journal} {\bibinfo  {journal} {Proc.
  R. Soc. London A}\ }\textbf {\bibinfo {volume} {241}},\ \bibinfo {pages}
  {376} (\bibinfo {year} {1957})}\BibitemShut {NoStop}%
\bibitem [{\citenamefont {Hansen}\ and\ \citenamefont
  {McDonald}(1986)}]{Hansen}%
  \BibitemOpen
  \bibfield  {author} {\bibinfo {author} {\bibfnamefont {J.-P.}\ \bibnamefont
  {Hansen}}\ and\ \bibinfo {author} {\bibfnamefont {I.~R.}\ \bibnamefont
  {McDonald}},\ }\href@noop {} {\emph {\bibinfo {title} {Theory of Simple
  Liquids}}},\ \bibinfo {edition} {2nd}\ ed.\ (\bibinfo  {publisher} {Academic
  Press},\ \bibinfo {address} {London},\ \bibinfo {year} {1986})\BibitemShut
  {NoStop}%
\bibitem [{\citenamefont {{Irving}}\ and\ \citenamefont
  {{Kirkwood}}(1950)}]{IrvingKirkwood}%
  \BibitemOpen
  \bibfield  {author} {\bibinfo {author} {\bibfnamefont {J.~H.}\ \bibnamefont
  {{Irving}}}\ and\ \bibinfo {author} {\bibfnamefont {J.~G.}\ \bibnamefont
  {{Kirkwood}}},\ }\bibfield  {title} {\enquote {\bibinfo {title} {{The
  Statistical Mechanical Theory of Transport Processes. IV. The Equations of
  Hydrodynamics}},}\ }\href@noop {} {\bibfield  {journal} {\bibinfo  {journal}
  {J. Chem. Phys.}\ }\textbf {\bibinfo {volume} {18}},\ \bibinfo {pages} {817}
  (\bibinfo {year} {1950})}\BibitemShut {NoStop}%
\bibitem [{\citenamefont {Ladd}\ \emph {et~al.}(1988)\citenamefont {Ladd},
  \citenamefont {Alley},\ and\ \citenamefont {Alder}}]{Ladd1988}%
  \BibitemOpen
  \bibfield  {author} {\bibinfo {author} {\bibfnamefont {A.~J.~C.}\
  \bibnamefont {Ladd}}, \bibinfo {author} {\bibfnamefont {W.~E.}\ \bibnamefont
  {Alley}}, \ and\ \bibinfo {author} {\bibfnamefont {B.~J.}\ \bibnamefont
  {Alder}},\ }\bibfield  {title} {\enquote {\bibinfo {title} {Shear viscosity
  and structural relaxation in dense liquids},}\ }\href@noop {} {\bibfield
  {journal} {\bibinfo  {journal} {Z. Phys. Chem}\ }\textbf {\bibinfo {volume}
  {156}},\ \bibinfo {pages} {331} (\bibinfo {year} {1988})}\BibitemShut
  {NoStop}%
\bibitem [{\citenamefont {Maier}\ \emph {et~al.}(2017)\citenamefont {Maier},
  \citenamefont {Zippelius},\ and\ \citenamefont {Fuchs}}]{Maier2017}%
  \BibitemOpen
  \bibfield  {author} {\bibinfo {author} {\bibfnamefont {M.}~\bibnamefont
  {Maier}}, \bibinfo {author} {\bibfnamefont {A.}~\bibnamefont {Zippelius}}, \
  and\ \bibinfo {author} {\bibfnamefont {M.}~\bibnamefont {Fuchs}},\ }\bibfield
   {title} {\enquote {\bibinfo {title} {Stress auto-correlation tensor in
  glass-forming isothermal fluids},}\ }\href@noop {} {\bibfield  {journal}
  {\bibinfo  {journal} {in preparation}\ } (\bibinfo {year}
  {2017})}\BibitemShut {NoStop}%
\bibitem [{\citenamefont {{Kadanoff}}\ and\ \citenamefont
  {{Martin}}(1963)}]{Kadanoff}%
  \BibitemOpen
  \bibfield  {author} {\bibinfo {author} {\bibfnamefont {L.~P.}\ \bibnamefont
  {{Kadanoff}}}\ and\ \bibinfo {author} {\bibfnamefont {P.~C.}\ \bibnamefont
  {{Martin}}},\ }\bibfield  {title} {\enquote {\bibinfo {title} {{Hydrodynamic
  equations and correlation functions}},}\ }\href@noop {} {\bibfield  {journal}
  {\bibinfo  {journal} {Annals of Physics}\ }\textbf {\bibinfo {volume} {24}},\
  \bibinfo {pages} {419} (\bibinfo {year} {1963})}\BibitemShut {NoStop}%
\bibitem [{\citenamefont {G\"otze}\ and\ \citenamefont {Latz}(1989)}]{Latz}%
  \BibitemOpen
  \bibfield  {author} {\bibinfo {author} {\bibfnamefont {W.}~\bibnamefont
  {G\"otze}}\ and\ \bibinfo {author} {\bibfnamefont {A.}~\bibnamefont {Latz}},\
  }\bibfield  {title} {\enquote {\bibinfo {title} {Generalised constitutive
  equations for glassy systems},}\ }\href@noop {} {\bibfield  {journal}
  {\bibinfo  {journal} {J. Phys. Condens. Matter}\ }\textbf {\bibinfo {volume}
  {1}},\ \bibinfo {pages} {4169} (\bibinfo {year} {1989})}\BibitemShut
  {NoStop}%
\bibitem [{\citenamefont {Evans}(1981)}]{Evans1981}%
  \BibitemOpen
  \bibfield  {author} {\bibinfo {author} {\bibfnamefont {D.}~\bibnamefont
  {Evans}},\ }\bibfield  {title} {\enquote {\bibinfo {title} {Equilibrium
  fluctuation expressions for the wave-vector- and frequency-dependent shear
  viscosity},}\ }\href@noop {} {\bibfield  {journal} {\bibinfo  {journal}
  {Phys. Rev. A}\ }\textbf {\bibinfo {volume} {23}},\ \bibinfo {pages} {2622}
  (\bibinfo {year} {1981})}\BibitemShut {NoStop}%
\bibitem [{\citenamefont {G\"{o}tze}(2009)}]{Goetze}%
  \BibitemOpen
  \bibfield  {author} {\bibinfo {author} {\bibfnamefont {W.}~\bibnamefont
  {G\"{o}tze}},\ }\href@noop {} {\emph {\bibinfo {title} {Complex Dynamics of
  Glass-Forming Liquids, A Mode-Coupling Theory}}}\ (\bibinfo  {publisher}
  {Oxford University Press},\ \bibinfo {year} {2009})\BibitemShut {NoStop}%
\bibitem [{\citenamefont {Leutheusser}(1984)}]{Leutheusser}%
  \BibitemOpen
  \bibfield  {author} {\bibinfo {author} {\bibfnamefont {E.}~\bibnamefont
  {Leutheusser}},\ }\bibfield  {title} {\enquote {\bibinfo {title} {Dynamical
  model of the liquid-glass transition},}\ }\href@noop {} {\bibfield  {journal}
  {\bibinfo  {journal} {Phys. Rev. A}\ }\textbf {\bibinfo {volume} {29}},\
  \bibinfo {pages} {2765} (\bibinfo {year} {1984})}\BibitemShut {NoStop}%
\bibitem [{\citenamefont {Desmond}\ and\ \citenamefont
  {Weeks}(2015)}]{Desmond2015}%
  \BibitemOpen
  \bibfield  {author} {\bibinfo {author} {\bibfnamefont {K.~W.}\ \bibnamefont
  {Desmond}}\ and\ \bibinfo {author} {\bibfnamefont {E.~R.}\ \bibnamefont
  {Weeks}},\ }\bibfield  {title} {\enquote {\bibinfo {title} {Measurement of
  stress redistribution in flowing emulsions},}\ }\href@noop {} {\bibfield
  {journal} {\bibinfo  {journal} {Phys. Rev. Lett.}\ }\textbf {\bibinfo
  {volume} {115}},\ \bibinfo {pages} {098302} (\bibinfo {year}
  {2015})}\BibitemShut {NoStop}%
\bibitem [{\citenamefont {Ahluwalia}\ and\ \citenamefont
  {Das}(1998)}]{Ahluwalia1998}%
  \BibitemOpen
  \bibfield  {author} {\bibinfo {author} {\bibfnamefont {R.}~\bibnamefont
  {Ahluwalia}}\ and\ \bibinfo {author} {\bibfnamefont {S.P.}\ \bibnamefont
  {Das}},\ }\bibfield  {title} {\enquote {\bibinfo {title} {Growing length
  scale related to the solidlike behavior in a supercooled liquid},}\
  }\href@noop {} {\bibfield  {journal} {\bibinfo  {journal} {Phys. Rev. E}\
  }\textbf {\bibinfo {volume} {57}},\ \bibinfo {pages} {5771} (\bibinfo {year}
  {1998})}\BibitemShut {NoStop}%
\bibitem [{\citenamefont {Goldhirsch}\ and\ \citenamefont
  {Goldenberg}(2002)}]{Goldhirsch2002}%
  \BibitemOpen
  \bibfield  {author} {\bibinfo {author} {\bibfnamefont {I.}~\bibnamefont
  {Goldhirsch}}\ and\ \bibinfo {author} {\bibfnamefont {C.}~\bibnamefont
  {Goldenberg}},\ }\bibfield  {title} {\enquote {\bibinfo {title} {On the
  microscopic foundations of elasticity},}\ }\href@noop {} {\bibfield
  {journal} {\bibinfo  {journal} {Eur. Phys. Jour. E}\ }\textbf {\bibinfo
  {volume} {9}},\ \bibinfo {pages} {245} (\bibinfo {year} {2002})}\BibitemShut
  {NoStop}%
\bibitem [{\citenamefont {Mukhopadhyay}\ \emph {et~al.}(2004)\citenamefont
  {Mukhopadhyay}, \citenamefont {Goldbart},\ and\ \citenamefont
  {Zippelius}}]{Mukhopadahyay2004}%
  \BibitemOpen
  \bibfield  {author} {\bibinfo {author} {\bibfnamefont {S.}~\bibnamefont
  {Mukhopadhyay}}, \bibinfo {author} {\bibfnamefont {P.~M.}\ \bibnamefont
  {Goldbart}}, \ and\ \bibinfo {author} {\bibfnamefont {A.}~\bibnamefont
  {Zippelius}},\ }\bibfield  {title} {\enquote {\bibinfo {title} {Goldstone
  fluctuations in the amorphous solid state},}\ }\href@noop {} {\bibfield
  {journal} {\bibinfo  {journal} {EPL (Europhysics Letters)}\ }\textbf
  {\bibinfo {volume} {67}},\ \bibinfo {pages} {49} (\bibinfo {year}
  {2004})}\BibitemShut {NoStop}%
\bibitem [{\citenamefont {Furukawa}\ and\ \citenamefont
  {Tanaka}(2011)}]{Furukawa2011}%
  \BibitemOpen
  \bibfield  {author} {\bibinfo {author} {\bibfnamefont {A.}~\bibnamefont
  {Furukawa}}\ and\ \bibinfo {author} {\bibfnamefont {H.}~\bibnamefont
  {Tanaka}},\ }\bibfield  {title} {\enquote {\bibinfo {title} {Direct evidence
  of heterogeneous mechanical relaxation in supercooled liquids},}\ }\href@noop
  {} {\bibfield  {journal} {\bibinfo  {journal} {Phys. Rev. E}\ }\textbf
  {\bibinfo {volume} {84}},\ \bibinfo {pages} {061503} (\bibinfo {year}
  {2011})}\BibitemShut {NoStop}%
\bibitem [{\citenamefont {Puscasu}\ \emph {et~al.}(2010)\citenamefont
  {Puscasu}, \citenamefont {Todd}, \citenamefont {Daivis},\ and\ \citenamefont
  {Hansen}}]{Puscasu2010}%
  \BibitemOpen
  \bibfield  {author} {\bibinfo {author} {\bibfnamefont {R.~M.}\ \bibnamefont
  {Puscasu}}, \bibinfo {author} {\bibfnamefont {B.~D.}\ \bibnamefont {Todd}},
  \bibinfo {author} {\bibfnamefont {P.~J.}\ \bibnamefont {Daivis}}, \ and\
  \bibinfo {author} {\bibfnamefont {J.~S.}\ \bibnamefont {Hansen}},\ }\bibfield
   {title} {\enquote {\bibinfo {title} {Nonlocal viscosity of polymer melts
  approaching their glassy state},}\ }\href@noop {} {\bibfield  {journal}
  {\bibinfo  {journal} {J. Chem. Phys.}\ }\textbf {\bibinfo {volume} {133}},\
  \bibinfo {pages} {144907} (\bibinfo {year} {2010})}\BibitemShut {NoStop}%
\bibitem [{\citenamefont {Olmsted}(2008)}]{Olmsted2008}%
  \BibitemOpen
  \bibfield  {author} {\bibinfo {author} {\bibfnamefont {Peter~D.}\
  \bibnamefont {Olmsted}},\ }\bibfield  {title} {\enquote {\bibinfo {title}
  {Perspectives on shear banding in complex fluids},}\ }\href {\doibase
  10.1007/s00397-008-0260-9} {\bibfield  {journal} {\bibinfo  {journal}
  {Rheologica Acta}\ }\textbf {\bibinfo {volume} {47}},\ \bibinfo {pages}
  {283--300} (\bibinfo {year} {2008})}\BibitemShut {NoStop}%
\bibitem [{\citenamefont {Picard}\ \emph {et~al.}(2002)\citenamefont {Picard},
  \citenamefont {Ajdari}, \citenamefont {Bocquet},\ and\ \citenamefont
  {Lequeux}}]{Picard2002}%
  \BibitemOpen
  \bibfield  {author} {\bibinfo {author} {\bibfnamefont {G.}~\bibnamefont
  {Picard}}, \bibinfo {author} {\bibfnamefont {A.}~\bibnamefont {Ajdari}},
  \bibinfo {author} {\bibfnamefont {L.}~\bibnamefont {Bocquet}}, \ and\
  \bibinfo {author} {\bibfnamefont {F.}~\bibnamefont {Lequeux}},\ }\bibfield
  {title} {\enquote {\bibinfo {title} {Simple model for heterogeneous flows of
  yield stress fluids},}\ }\href@noop {} {\bibfield  {journal} {\bibinfo
  {journal} {Phys. Rev. E}\ }\textbf {\bibinfo {volume} {66}},\ \bibinfo
  {pages} {051501} (\bibinfo {year} {2002})}\BibitemShut {NoStop}%
\bibitem [{\citenamefont {Goyon}\ \emph {et~al.}(2010)\citenamefont {Goyon},
  \citenamefont {Colin},\ and\ \citenamefont {Bocquet}}]{Goyon2010}%
  \BibitemOpen
  \bibfield  {author} {\bibinfo {author} {\bibfnamefont {J.}~\bibnamefont
  {Goyon}}, \bibinfo {author} {\bibfnamefont {A.}~\bibnamefont {Colin}}, \ and\
  \bibinfo {author} {\bibfnamefont {L.}~\bibnamefont {Bocquet}},\ }\bibfield
  {title} {\enquote {\bibinfo {title} {How does a soft glassy material flow:
  finite size effects{,} non local rheology{,} and flow cooperativity},}\
  }\href@noop {} {\bibfield  {journal} {\bibinfo  {journal} {Soft Matter}\
  }\textbf {\bibinfo {volume} {6}},\ \bibinfo {pages} {2668--2678} (\bibinfo
  {year} {2010})}\BibitemShut {NoStop}%
\bibitem [{\citenamefont {Bouzid}\ \emph {et~al.}(2015)\citenamefont {Bouzid},
  \citenamefont {Izzet}, \citenamefont {Trulsson}, \citenamefont {Cl\'ement},
  \citenamefont {Claudin},\ and\ \citenamefont {Andreotti}}]{Bouzid2015}%
  \BibitemOpen
  \bibfield  {author} {\bibinfo {author} {\bibfnamefont {M.}~\bibnamefont
  {Bouzid}}, \bibinfo {author} {\bibfnamefont {A.}~\bibnamefont {Izzet}},
  \bibinfo {author} {\bibfnamefont {M.}~\bibnamefont {Trulsson}}, \bibinfo
  {author} {\bibfnamefont {E.}~\bibnamefont {Cl\'ement}}, \bibinfo {author}
  {\bibfnamefont {P.}~\bibnamefont {Claudin}}, \ and\ \bibinfo {author}
  {\bibfnamefont {B.}~\bibnamefont {Andreotti}},\ }\bibfield  {title} {\enquote
  {\bibinfo {title} {Non-local rheology in dense granular flows - revisiting
  the concept of fluidity},}\ }\href@noop {} {\bibfield  {journal} {\bibinfo
  {journal} {Eur. Phys. J. E}\ }\textbf {\bibinfo {volume} {38}},\ \bibinfo
  {pages} {125} (\bibinfo {year} {2015})}\BibitemShut {NoStop}%
\end{thebibliography}%


\end{document}